# Fully Quantum Algorithm for the 1-dimensional linear Lattice Boltzmann Method

Mohammed Bediche [1,2,3] (✉)[0009-0001-0255-3894], Matthijs van Waveren [1][0000-0001-7280-9114], Denis Ricot[2], and Pierre Sagaut [3][0000-0002-3785-120X]

[1] CS Research Lab, 3, boulevard Thomas Gobert, 91120 Palaiseau, France
mohammed.bediche@cs-soprasteria.com,
matthijs.vanwaveren@compilaflows.com
[2] CS GROUP, 22, avenue Galilée, 92350 Le Plessis-Robinson, France
denis.ricot@cs-soprasteria.com
[3] Aix Marseille Univ, Centrale Med, CNRS, M2P2 Laboratory, 13013 Marseille, France
pierre.sagaut@univ-amu.fr

**Abstract.** A fully quantum algorithm for solving the one-dimensional linear advection-diffusion equation using the Lattice Boltzmann method as a numerical procedure is presented in this work. We start by presenting a state of the art of the current usage of quantum algorithms for solving ordinary and partial differential equations. We then describe two algorithms for the one-dimensional Lattice Boltzmann method with two degrees of freedom. The first one is an existing hybrid quantum-classical algorithm with measurements at each time step, and the second one is our improved version, viz. a fully quantum algorithm where only one measurement is needed at the end of the algorithm. The fully quantum algorithm is first executed on a quantum simulator and then compared with a classical approach. Subsequently, the fully quantum algorithm is run on a quantum system with 133 qubits to investigate the effect of noise and the depth of the circuit on the output state. We find fluctuations in the final result due to the decoherence noise of the qubits.

## 1 Introduction

Differential equations constitute the foundational framework for the mathematical modeling of physical systems. However, executing numerical simulations of nonlinear ordinary and partial differential equations (ODEs/PDEs) on classical computers remains a formidable challenge, primarily due to the extensive computational time and data processing resources required. In response to these limitations, quantum programming has emerged as a promising alternative, garnering increasing attention over recent decades. Certain computational problems that are intractable for classical machines can be addressed theoretically using quantum computers, for example, Shor's factoring algorithm, which underpins the concept of quantum supremacy. A key advantage of quantum computing lies in its potential to exponentially expand the dimensionality of the state vector—hence the number of grid points in a simulation—as the number of qubits increases, thereby enabling the manipulation of high-dimensional state spaces. Quantum algorithms, owing to their inherently linear structure, are particularly well-suited for solving linear differential equations. Several such algorithms offer a speed-up over their classical counterparts [1,2]. Nevertheless, progress in developing quantum algorithms for nonlinear differential equations has been comparatively limited, and these approaches have yet to outperform established classical techniques. Broadly speaking, the numerical solution of PDEs can be approached via two principal methods. The first involves discretizing the system using integration techniques such as finite difference or finite element methods, implemented with either a HHL-based quantum algorithm [1,3] or a VQA algorithm [4]. The second strategy consists in transforming the problem into a Schrödinger-type equation, which can then be addressed using Hamiltonian simulation techniques [5].

In the field of fluid mechanics and the simulation of turbulent flows, the partial differential equations (PDEs) that are generally considered are the Navier–Stokes equations. For industrial flow configurations, simplified turbulence models must be employed within the system of equations, which leads to non-negligible discrepancies with respect to the actual physics. The potential capability of quantum simulation to address very large-scale models while maintaining the shortest possible computation times is of great interest for a wide range of physical domains, including electromagnetic computations, structural analysis, acoustics, materials, combustion, and many others. For fluid mechanics simulations, an alternative to solving the Navier–Stokes equations is the use of the Lattice Boltzmann Method (LBM) [6]. This numerical method enables the simulation of various convection–diffusion phenomena, both linear and nonlinear. Several academic and commercial software packages based on LBM are available, including ProLB, developed by CS GROUP within the framework of an industrial–academic partnership.

In this work, we present a quantum algorithm for the 1-dimensional linear advection-diffusion equation solved with the Lattice Boltzmann Method. We start by following the approach of L. Budinski [7], who wrote a hybrid quantum-classical algorithm that resembles the classical algorithm of the Lattice-Boltzmann Method. One of the drawbacks of this approach is the need of repeated measurements and reinitialization at each time step. This increases the complexity of the algorithm and hinders any potential for quantum advantage in the long run. We improve on the hybrid classical-quantum algorithm by proposing a fully quantum algorithm for the one-dimensional Lattice Boltzmann formulation D1Q2. We simulate the fully quantum algorithm on the Qiskit quantum simulator, and we execute it on the 133-qubit IBM quantum computer `ibm_torino`. The results from the computers have the same evolution of the flow as the classical results and the results from the quantum simulator. The results from the quantum computer however exhibit fluctuations due to the decoherence noise in the qubits.

## 2 Lattice Boltzmann Method

The advection-diffusion equation solved with the Lattice Boltzmann Method is the subject of this paper.

$$\frac{\partial \phi}{\partial t} + \frac{\partial (u_i \phi)}{\partial x_i} = \frac{\partial}{\partial x_i}\left(D \frac{\partial \phi}{\partial x_i}\right) \tag{1}$$

where $\phi$ is the depended variable (mass, momentum, energy, species, etc.), $t$ is time, $x_i$ is a Cartesian coordinate, $u_i$ is the fluid velocity in the $i$-direction and $D$ is the diffusion coefficient. The above equation is valid for general advection and diffusion phenomena, including both steady and unsteady situations.

In the Lattice Boltzmann Method, instead of solving the nonlinear equation, we solve an equivalent problem with a higher dimension. However, this formulation still contains a non-linear collision term that is not easily implemented on a quantum computer. Previous works have either neglected the non-linear part of this operator [7,8] or proceeded by linearizing it using Carleman linearization [9]. The non-linear part is zero if the advection velocity is independent of the concentration and can be neglected if the mass transport is dominated by diffusion.

In the Lattice Boltzmann method, a particle density function $f(x, c, t)$ is used to describe the probability of finding one distribution of particles within the elemental cube (x,x+dx) with a velocity comprised within the interval (c,c+dc) at the time t. The discretized Boltzmann equation under the Bhatnagar-Gross-Krook (BGK) approximation in the phase space [10] is given by:

$$\frac{\partial f_\alpha(\vec{x},t)}{\partial t} + c_{\alpha,i}\frac{\partial f_\alpha(\vec{x},t)}{\partial x_i} = -\frac{f_\alpha(\vec{x},t) - {f^{eq}}_\alpha(\vec{x},t)}{\tau} \tag{2}$$

The index $i$ refers to the discretization in space and the index α defines the discretization in the velocity components. These distribution functions are linked to the macroscopic variables by:

$$\rho(\vec{x},t) = \sum_\alpha f_\alpha(\vec{x},t)\ ;\ \rho.\vec{u}(\vec{x},t) = \sum_\alpha \vec{c_\alpha}.f_\alpha(\vec{x},t) \tag{3}$$

The algorithm of the Lattice Boltzmann Method is a four-step process: i) initialization, ii) collision, iii) propagation and iv) macro-states calculation [11]. We develop these steps in the next section.

### 2.1 Fully Quantum Algorithm for the Linear D1Q2

We describe a fully quantum algorithm that uses the Lattice Boltzmann Method to solve the linear advection-diffusion equation in a one-dimensional space. This algorithm builds upon the work of Budinski [7] where he presented a hybrid classical-quantum algorithm for the linear D1Q2 model with a relaxation time equals to 1. In the D1Q2 model, the index α in (2) and (3) has values 1 and 2 with only two velocity components, $e_1 = -e_2 = \frac{\Delta x}{\Delta t}$. Additionally, we consider $\Delta x = \Delta t$. The Boltzmann equation becomes:

$$f_\alpha(x + e_\alpha \Delta t, t + \Delta t) = f^{eq}{}_\alpha \tag{4}$$

The linear equilibrium distribution functions are given by (5) with $\omega_\alpha = 0.5$:

$$f_\alpha{}^{eq}(\vec{x}, t) = \omega_\alpha \rho(\vec{x}, t)\big(1 + e_{\alpha,i} u_i\big) + o(u_i) = A_\alpha\, \rho(\vec{x}, t) \tag{5}$$

$A_\alpha$ are square diagonal matrices of size $N \times N$, with N the number of lattice points. We consider the lattice points $x_0, x_1, \ldots, x_{N-1}$ each describing a position of the distribution function on this lattice. To encode the model into our quantum algorithm, $log_2(N)$ qubits are needed with one additional qubit used to encode information about the two degrees of freedom. One ancilla qubit is added to aid calculations.

We encode the density at t=0 in the normalized state vector of eq (6). We applied the reverse iterative procedure proposed by Shende et al. [15], which is implemented in the built-in function `initialize` in Qiskit.

$$|\psi_0\rangle = |0\rangle_a \sum_{i=0}^{2N-1} \frac{\rho_i}{\|\psi\|} |i\rangle \tag{6}$$

The following circuit in Fig.1 shows the original hybrid quantum circuit published by Budinski [7]. It represents the 4 steps needed in the quantum Lattice Boltzmann algorithm. The dashed vertical lines separate the four sections of the algorithm. We first describe this circuit before describing our improvement to convert it to a fully quantum algorithm.

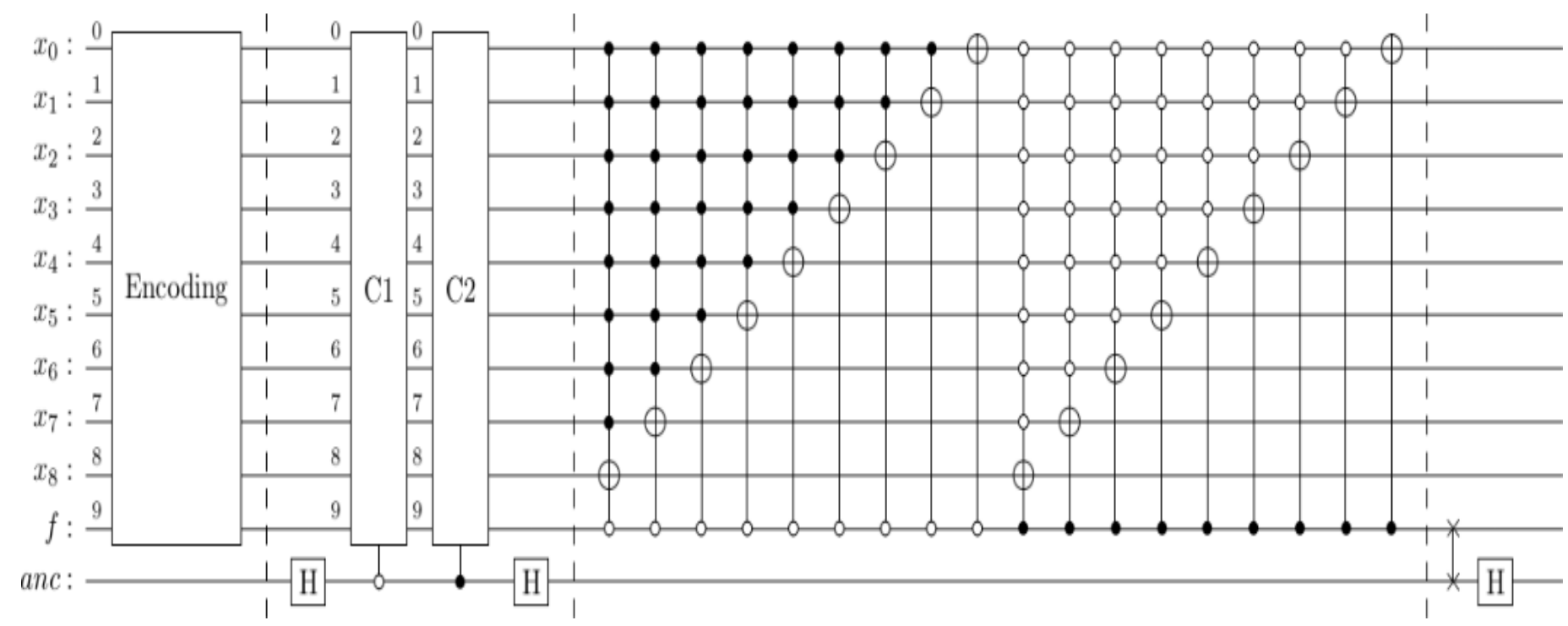

**Fig. 1.** The quantum circuit for one timestep. The leftmost section is the encoding step. The section with C1 and C2 is the collision step. The section to the right of it is the propagation step. The final rightmost section is the macro-states calculation step.

**Collision.** For the collision step, the Linear Combination of Unitaries approach [11] is used by applying the following set of operators:

$$|\psi_1\rangle = \left(H_a{}^{+}\otimes I_q\right)\left(|0\rangle\langle 0|_a C_1 + |1\rangle\langle 1|_a C_2\right)\left(H_a\otimes I_q\right)|\psi_0\rangle \tag{7}$$

where $C_1$ and $C_2$ are unitary diagonal operators constructed from the matrices $A_1$ and $A_2$ and are described in [7]. The state vector post-collision is:

$$|\psi_1\rangle = \frac{1}{\|\psi\|}\left(|0\rangle_a \sum_{i=0}^{2N-1} f_{\alpha,i}|i\rangle + i|1\rangle_a \sqrt{I-A^2}\sum_{i=0}^{2N-1}\psi_{\alpha,i}|i\rangle\right) \tag{8}$$

**Propagation.** The propagation step is carried out using the quantum walk procedure, taking advantage of the fact that the quantum state is already in superposition. This is done using a series of multiple-controlled NOT, a C-NOT gate and a NOT gate. The difference between the two directions of propagation is the state of the control qubit. These operators can be seen in Fig 1. The state vector $|\psi_2\rangle$ after propagation is:

$$|\psi_2\rangle = \hat{L}\hat{R}|\psi_1\rangle \tag{9}$$

**Macro-states calculations.** To calculate the particle density and the velocity, we first followed Budinski's algorithm and applied a SWAP gate between the ancilla qubit and the N-1'th qubit followed by a Hadamard. The final state is:

$$\begin{aligned}|\psi_3\rangle = \frac{1}{\sqrt{(2)}\|\psi\|}\Bigg(&|0\rangle_a|0\rangle_{N-1}\sum_{i=0}^{N-1}(f_1+f_2)|i\rangle + |1\rangle_a|0\rangle_{N-1}\sum_{i=0}^{N-1}(f_1-f_2)|i\rangle \\ &+ |0\rangle_a|1\rangle_{N-1}(C1-C2)\sum_{i=0}^{N-1}\psi_{\alpha,i}|i\rangle\Bigg)\end{aligned} \tag{10}$$

In the hybrid scheme, the resulting state vector is measured, and classical post-processing is performed to extract the density and the velocity. These are then input in the calculation of $|\psi_0\rangle$ of the next timestep. Each timestep requires thus executing a quantum circuit, measuring the resulting state vector, and performing classical post-processing to extract the density and the velocity. This procedure incurs significant overhead. For bigger systems where the Hilbert space has a large dimension, the algorithm behaves worse than a classical algorithm. A fully quantum algorithm is needed to exploit the full potential of quantum machines.

**Fully quantum algorithm.** To build a quantum algorithm that avoids repeated measurements, we start by the quantum state $|\psi_3\rangle$ calculated previously. The state vector shows that the information about $f_1 + f_2$ and $f_1 - f_2$ is already stored in the final state vector just before the measurements. We now develop the expression of the equilibrium distribution function and show that it contains these two quantities. It is easy to show that:

$$\rho(\vec{x}, t + \Delta t) = f_1(\vec{x}, t + \Delta t) + f_2(\vec{x}, t + \Delta t)\ ;$$
$$\rho.\overrightarrow{u}(\vec{x}, t + \Delta t) = f_1(\vec{x}, t + \Delta t)\ -\ f_2(\vec{x}, t + \Delta t) \tag{11}$$

From (5) and (11), we derive the following expressions:

$$f^{eq}{}_1(x, t) = \frac{1}{2}\big(\rho(\vec{x}, t) +\ \rho.\overrightarrow{u}(\vec{x}, t)\big);$$
$$f^{eq}{}_2(x, t) = \frac{1}{2}\big(\rho(\vec{x}, t) -\ \rho.\overrightarrow{u}(\vec{x}, t)\big) \tag{12}$$

This means that it is as if we had the result of the collision at the time step *t+1* encoded in the state obtained after the calculations at the instant *t*. We just need to rearrange the states again to separate these two values. To do that, we modify the circuit of Fig. 1 by adding a SWAP gate between the sixth qubit in the *q*-register and the ancilla qubits and apply a Hadamard gate on qubit *N-1*. We obtain the state in equation (13). $|\lambda\rangle_q$ is an additional state that holds no useful information to us.

$$|\psi_3\rangle = \frac{1}{\|\psi\|}\left(|0\rangle_a|0\rangle_{N-1}\sum_{i=0}^{N-1} f^{eq}{}_{1,i}|i\rangle + |0\rangle_a|1\rangle_{N-1}\sum_{i=0}^{N-1} f^{eq}{}_{2,i}|i\rangle + |1\rangle_a|\lambda\rangle_q\right) \tag{13}$$

The state $|1\rangle_a|\lambda\rangle_q$ is for the other amplitudes obtained that has no exploitable information in this algorithm which we didn't write for simplicity. We see that we do not need to switch to a classical algorithm to prepare the system for future time steps, and that we can stay in the quantum space. Consequently, we can directly follow up by the propagation and the macro-states calculations, which drastically optimizes the time needed for the algorithm. The overall architecture of the algorithm is given in Fig 2 where $|\psi_4\rangle$ is the state vector obtained after the quantum post-processing steps which allows us to bypass the intermediate measurement steps and classical processing in the hybrid algorithm to execute the next run.

Initialisation $\rho(x, t_0) \rightarrow |\psi_0\rangle$ → Collision $|\psi_1\rangle$ → Propagation $|\psi_2\rangle$ → Macro-variables $|\psi_3\rangle$

$SWAP_{a-q_{n-1}} \otimes H_{q_n-1}|\psi_3\rangle \rightarrow |\psi_4\rangle$

**Fig. 2.** The architecture of the quantum algorithm. The black arrows represent the first run of the algorithm and the grey ones the additional post treatment and the processes repeated in the *m* time loops afterwards for a total time of *t=mΔt*.

Our method eliminates the need for intermediate classical feedback. Instead, the required updates are achieved entirely within the quantum circuit by appending only two additional quantum gates at the end of each run. This results in a more efficient computation.

**Complexity** To calculate the complexity of the quantum algorithm in Fig. 2, we first divide it into two parts: (a) a first forward part with the four major steps: encoding, collision, propagation and macro-states calculation and (b) a second loop part containing the quantum post-processing, the propagation and the calculation of the macroscopic variables. This second part is the one which is looped over once for each timestep.

In the first forward part, encoding the input variables into the state vector requires $2 \times 4^n - (2n + 3) \times 2^n + 2n$ CNOT gates with *n* being the number of qubits according to Table 1 in [15]. For the collision step, we require $O(\log_2 m)$ ancilla qubits for *m* directions of propagation. The LCU scales as $O(1)$ in *n*, since the number of unitaries does not grow with the system size. The multi-controlled NOT gates used in the propagation step scale as $O(n^2)$ in the number of CNOT gates [16]. The macro-states calculations require 3 CNOT gates and one Hadamard gate so its complexity is negligible.

For the second loop part of our algorithm, we require $O(t(n)^2)$ CNOT gates with $t$ being the number of timesteps. We see that the depth of the second part of the circuit in terms of CNOTs is linear in $t$ and quadratic in *n*. Note that the number of qubits $n = \log_2 N + \log_2 m$, with *N* the number of grid points and *m* the number of directions of propagation.

# 3 Results

## 3.1 Execution on a Quantum Simulator

We ran the simulations using a centered Gaussian flow as input with periodic boundary conditions. The domain is discretized using 512 regular mesh points with 1000-time steps. The encoding of the initial values of the concentration was done using the instruction `initialize` available on Qiskit, which encodes an arbitrary quantum gate into the register of n qubits. The results of the fully quantum algorithm matched the ones obtained using the classical one, as shown in Fig 3. Two time-instances, t=200 and t=600, were considered to show the evolution of the flow.

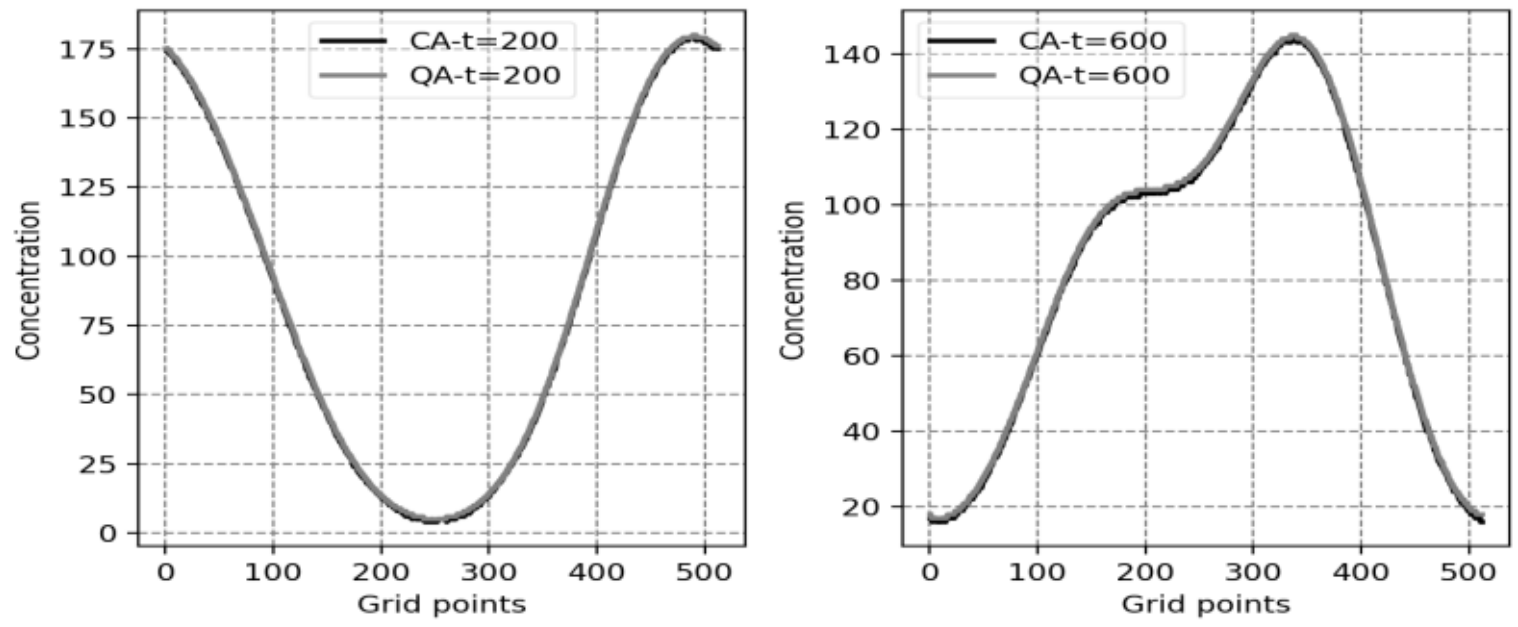

**Fig. 3.** Results of the fully quantum algorithm on a quantum simulator. CA corresponds to the Classical Algorithm, and QA to the Quantum Algorithm.

The original hybrid version of the algorithm was also compared with the classical one giving matching results. However, since the hybrid algorithm requires several measurement operations for each time step, it requires a longer time to execute.

### 3.2 Performance of the Fully Quantum Algorithm on a Quantum Computer

The fully quantum algorithm was then tested on the 133-qubit quantum machine `ibm_torino`. For every gate present in the quantum algorithm, a decomposition step is necessary to implement the gate on the quantum machine. For `ibm_torino`, there are 7 elementary gates which are the CZ, Id, RX, RZ, RZZ, SX and X. Every customized gate in the algorithm needs to be decomposed into these elementary gates before the algorithm is executed on the machine.

The quantum computer returns a histogram, which is the probability distribution obtained following the measurement operations at the end of the algorithm. Hence, we cannot directly access the quantum state. However, since the useful information of the quantum state only has a real part, it is possible to reconstruct the profile of the density as a function of the mesh points from the probability distribution. Due to the presence of noise on the quantum hardware, it was deemed necessary to reduce the size of the systems to 8 mesh points in order to decrease the depth of the algorithm. Multiple timesteps were also computed but as expected the output became noisier the longer the simulation.

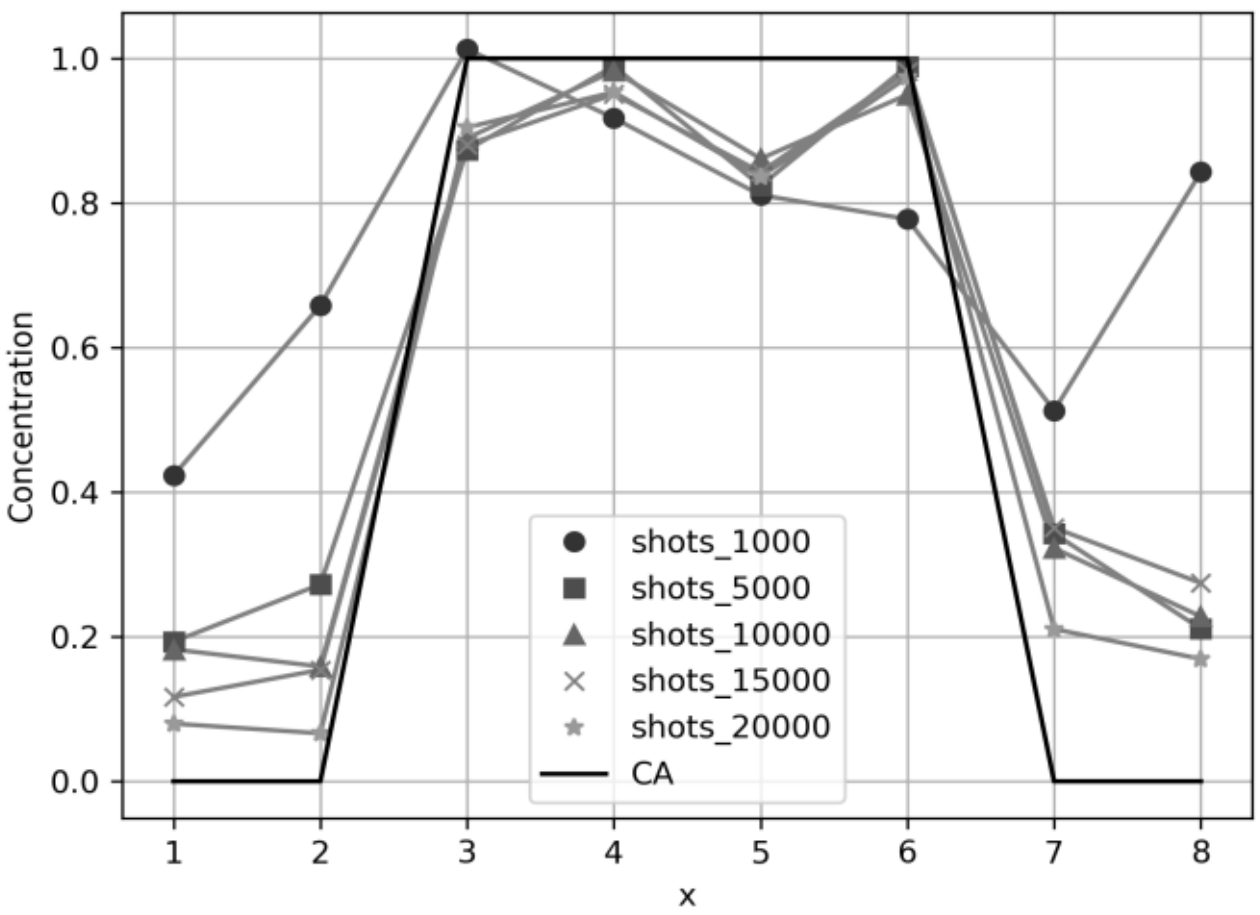

**Fig. 4.** Concentration results obtained from `ibm_torino`. The horizontal axis has the meshpoints.

Fig.4 represents the result obtained from `ibm_torino`, and it shows that we were successful in obtaining the same evolution of the flow as the results on the classical algorithm. We also note that increasing the number of shots from 1000 to 20000 led to better results. The values of the concentration from the run on the quantum machine fluctuate and differ from the values of the classical ones, but they have the same qualitative evolution of the flow. Our hypothesis is that this difference is due to the decoherence noise in the qubits, because larger systems and longer simulations lead to more noise. The noise in the transmon qubits arises mainly from the effects of dielectric loss and its manifestation through the interaction with two-level systems (TLS) [13]. To overcome the problem of noise in the qubits, quantum error mitigation and quantum error correction techniques are utilized. The basic idea of quantum error correction is the introduction of redundancy in the qubits, which means that one logical error-free qubit is represented by dozens of physical noisy qubits. Brayyi et al [14] for example present an end-to-end quantum error correction protocol that implements fault-tolerant memory on the basis of a family of low-density parity-check codes. We expect that once the error mitigation and correction techniques have been implemented in production machines, we will be able to simulate larger systems.

## 4 Discussion

The goal of this work was to show that we can build a fully quantum algorithm for the one-dimensional linear advection-diffusion equation using the lattice Boltzmann method as a numerical procedure. The focus was more on avoiding measurements as much as possible and not as much on optimizing the steps and the gates in the algorithm. We were inspired by the hybrid quantum-classical algorithm of L. Budinski [7]. This hybrid quantum-classical algorithm is limited in its scope for quantum speedup due to the classical measurement bottleneck. We developed a fully quantum version of the algorithm by replacing the measurement step with a state vector rearrangement step. We tested this algorithm both on the Qiskit quantum simulator and on the 133-qubit IBM quantum computer `ibm_torino`. The results from the quantum computer have the same evolution of the flow as the results from the quantum simulator. The results from the quantum computer exhibit fluctuations due to the noise in the qubits. Our future work is to reduce these fluctuations by decreasing the depth of the circuits by optimizing the encoding, collision and propagation steps. We are also working on extending this algorithm to 2D systems and adding non-linearity to the model.

## Appendix

The quantum algorithm was implemented using a Jupyter notebook. The version of Python that was used is 3.12.7. The API Qiskit version 1.2.4 was used with:

- Qiskit-aer: version 0.15.1
- Qiskit-ibm-runtime: version 0.30.0

The initial density is a centered Gaussian distribution, and the initial uniform velocity field is 0.2. The code is not open source, thus we cannot publish it on GitHub.

**Acknowledgements.** This study was partly funded by the French state through CIFRE.

**Disclosure of Interests.** The authors have no competing interests to declare that are relevant to the content of this article.